# PRINS: Resistive CAM Processing in Storage

Leonid Yavits, Roman Kaplan and Ran Ginosar

**ABSTRACT**

Near-data in-storage processing research has been gaining momentum in recent years. Typical processing-in-storage architecture places a single or several processing cores inside the storage and allows data processing without transferring it to the host CPU. Since this approach replicates von Neumann architecture inside storage, it is exposed to the problems faced by von Neumann architecture, especially the bandwidth wall. We present PRINS, a novel in-data processing-in-storage architecture based on Resistive Content Addressable Memory (RCAM). PRINS functions simultaneously as a storage and a massively parallel associative processor. PRINS alleviates the bandwidth wall faced by conventional processing-in-storage architectures by keeping the computing inside the storage arrays, thus implementing in-data, rather than near-data, processing. We show that PRINS may outperform a reference computer architecture with a bandwidth-limited external storage. The performance of PRINS Euclidean distance, dot product and histogram implementation exceeds the attainable performance of a reference architecture by up to four orders of magnitude, depending on the dataset size. The performance of PRINS SpMV may exceed the attainable performance of such reference architecture by more than two orders of magnitude.

## 1. INTRODUCTION

The premise of data centric, or near-data processing is reducing the memory access time by cutting the physical distance and increasing the bandwidth between CPU and memory. Since inception, data centric processing mainly meant processing in memory (PIM). To process datasets larger than memory footprint, processing moves further down the memory hierarchy, achieving processing-in-storage.

We believe that near-data processing-in-storage is inherently limited because it is largely based on replicating the von Neumann architecture in storage. Hence it potentially faces some of von Neumann architecture problems, such as the bandwidth wall.

This work presents PRINS, a novel resistive CAM (RCAM)-based *in-data* (rather than near-data) processing-in-storage architecture. PRINS simultaneously functions as data storage and a massively parallel SIMD accelerator that performs the computations *in-situ*, resulting in increased performance by better utilization of the internal storage bandwidth, and reduced energy consumption.

PRINS can be placed in different levels of the memory hierarchy. While the entire mass storage can be RCAM based, an intermediate storage level (located between the main memory and the mass storage), for example a storage class memory, could be a better cost/performance tradeoff.

This paper makes the following contributions:

- We introduce PRINS, a RCAM in-data processing-in-storage architecture. We present the bottom-top design, from the memristor-based RCAM bitcell, to the PRINS system and its position within computer memory hierarchy. We describe the programming model of PRINS and discuss its application interface.
- We develop PRINS based implementation of several microbenchmarks and algorithms in the fields of machine learning, data analytics and graph processing.
- We show that PRINS may significantly outperform a computer architecture with a bandwidth limited external storage while providing high power efficiency.

The rest of this paper is organized as follows. Section 2 presents the motivation and related work. Section 3 introduces the architecture of PRINS. Section 4 reviews the principles of associative processing. Section 5 explores PRINS programming and applications. Section 6 presents the evaluation. Section 7 offers conclusions.

## 2. Background and Motivation

Table 1 categorizes related work according to processing-in-memory vs. storage and near-data vs. in-data processing. Processing-in-memory includes cache and DRAM whereas storage relates mostly to non-volatile memory and storage. In near-data processing systems, processor cores are placed close to the data, whereas in-data processing refers to compute circuitry attached to each memory or storage bit.

A comprehensive review of data centric processing architectures and trends can be found in [9].

- *Leonid Yavits, E-mail: yavits@technion.ac.il.*
- *Roman Kaplan, E-mail: sromanka@campus.technion.ac.il*
- *Ran Ginosar, E-mail: ran@ee.technion.ac.il.*

*Authors are with the Department of Electrical Engineering, Technion-Israel Institute of Technology, Haifa 3200000, Israel.*
*Manuscript submitted:*

**Table 1. Data centric processing taxonomy**

| | Near-Data Processing | In-Data Processing |
|---|---|---|
| Processing-in-Memory | STARAN, DAPP, MPP [61]<br>Terasys [27]<br>DIVA [32]<br>HTMT [43]<br>PIM Lite [12]<br>Cyclops [6]<br>Gilgamesh [68]<br>SLIIC QL [69]<br>FlexRAM [40]<br>Computational RAM [22]<br>Tesseract [1]<br>Active Memory Cube [57][70]<br>Smart Memory Cube [7]<br>HAMLeT [3]<br>NDA [24] | DAAM [48]<br>ReTCAM [29][30]<br>AC-DIMM [31]<br>AP [78]<br>ReAP [77]<br>GP-SIMD [54]<br>Re-GP-SIMD [55]<br>PRIME [13]<br>ISAAC [66]<br>MAGIC [64][56]<br>MBARC [60]<br>APIM [37]<br>CIM [33]<br>PINATUBO [47]<br>MPIM [36]<br>Memristive Boltzmann machine [11] |
| Processing-in-Storage | Active Flash [10]<br>Intelligent SSD [8]<br>Smart SSD [41]<br>Collaborative in-SSD processing [38]<br>XSD [14]<br>Active Disk/iSSD [15]<br>Minerva [19]<br>BlueDBM [39] | PRINS (this work) |

## 2.1 Near-data processing-in-memory

While processing-in-storage research is relatively young, near-data processing-in-memory (PIM) has been thoroughly researched. The concept of mixing memory and logic has been around since 1960s [61][27][32][6][68][43][69][12].

While embedding processing with conventional 2D DRAM chips is less practical, recent advancement in 3D memory and logic stacking technology removed this obstacle. Recent years saw a variety of 3D memory/processing stack architectures [1][2][57][70][53][7][80] [3][24].

## 2.2 In-data processing-in-memory

In-data processing-in-memory developed in parallel with near-data processing-in-memory research [48][78][54]. Recently, emerging memory technologies such as resistive memory have become a focus of PIM research [60][13][77][55][66][64][33][37].

## 2.3 Near-data processing-in-storage

Typical processing-in-storage architecture places a single or several processing cores inside the storage (with main focus on NAND flash solid-state disk) and allows data processing without transferring it to the host processor [10][8][38][14][41][19][39][15]. The concept of near-data processing-in-storage is illustrated in Figure 1(a).

## 2.4 In-data processing-in-storage

We believe that near-data processing-in-storage is limited because it is based on replicating the von Neumann architecture in a storage. Hence it potentially faces some of the von Neumann architecture problems, such as the bandwidth wall. We define the computation throughput of an in-storage processor as follows:

$$Throughput = \frac{Dataset\_size\ [Byte]}{Runtime\ [sec]} \quad (1)$$

For processing-in-storage system to reach optimal performance, the peak computation throughput of an in-storage processor should be on par with the internal bandwidth of that storage. The upper bound of such bandwidth is defined by the maximum bandwidth of flash arrays, and ranges from few hundred MB/s to few GB/s depending on the number of parallel flash channels [59].

Earlier works on in-SSD processing report the computation throughput of several MB/s to a few hundred MB/s depending on workload (for example, 7MB/s to 350MB/s in [10]). However, as the number of flash channels in SSD grows, so does the effective internal SSD bandwidth. A conventional response to the growing internal bandwidth is increasing parallelism by adding more in-SSD processing cores. One example of such increased parallelism is placing a processing core in each flash channel [38]. However, with the advancement of non-charge based memory technologies, there is a growing consensus that resistive memory has a potential to replace flash in future SSDs [4].

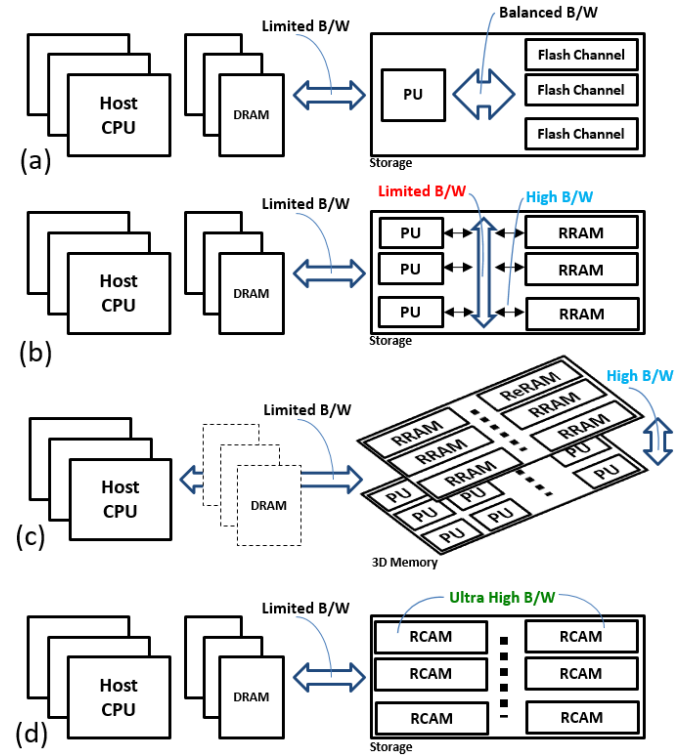

**Figure 1: (a) Near-Data Processing in Flash Based Storage; (b) Near-Data Processing in Resistive RAM (RRAM) Based Storage; (c) 3D Near-Data Processing-in-Memory; (d) In-data Processing-in-storage in PRINS.**

With bandwidth and latency characteristics similar to DRAM [16], resistive memory may significantly increase the upper bound of the internal SSD bandwidth. This may lead to the following two scenarios. First, increasing the parallelism by adding more in-SSD processing cores will become inefficient and may eventually cause a reduction in performance [76]. Second, internal storage bandwidth is likely to become limited by the internal communication bus/network (Figure 1(b) due to the surge in inter-core communication [76]. Both scenarios repeat the problems faced by the host multicore von Neumann architectures.

As suggested in [9], the computation throughput to internal SSD bandwidth balance can be regained through new system-on-chip and die stacking technologies that enable network-on-chip integration, a more efficient network software stack, and potentially new near-data processing-customized interconnect designs.

The concept of 3D near-data processing architecture is illustrated in Figure 1(c). The majority of 3D near-data processing research targets processing-in-memory. 3D stacking of DRAM and a multitude of processing cores allows significantly lifting the bandwidth limit (for example by 16× [1]). Similarly, 3D stacking of NVM storage with a parallel processor having advanced vertical communication capabilities may bring about the full realization of the bandwidth upside of RRAM. This is certainly a valid potential direction of the near-data processing-in-storage development.

In this paper, we propose a new in-data processing-in-storage architecture, PRINS, that increases the computation throughput to match the potentially ultra-high internal bandwidth of the storage arrays. This architecture progresses from random addressable to content addressable (associative) storage (Figure 1(d)). PRINS enables massively parallel SIMD processing of the data *inside* the storage arrays. The processing is associative, making conventional in-storage processors redundant. There is no data transfer outside the storage arrays through a bandwidth limited internal SSD communication bus/network. The inherent performance (read/write access time and bandwidth) of the resistive memory can be utilized to the full extent, enabling very high computation throughput while reducing the energy consumption (mainly due to the lack of data movement inside storage).

The main difference between PRINS and a hypothetical 3D near-data processing-in-storage architecture described above is the bitwise connectivity of memory and processing: In PRINS, each memory bit is directly connected to processing transistors, whereas in 3D near-data processing, the data must pass through memory interface circuits and through vertical interconnects, typically much fewer in numbers than the number of bits. In PRINS, the bulk of data ideally never leaves the memory. The computation is performed within the confines of the memory array. This potentially holds a significant performance and energy efficiency advantage: Using DRAM as example, there is typically a reduction in available bandwidth of six orders of magnitude between the sense amplifiers and the CPU edge [9]. In addition, the cost of access in terms of energy increases from hundreds of femtojoules to tens of picojoules over a span of the same distance [9].

The use of STT-MRAM and Resistive Ternary CAM for data intensive computing was pioneered by Guo *et al.* [29][30][31]. Guo *et al.* used the associative capabilities of CAM and Ternary CAM mainly for search operations, while the computing is largely done in a CPU. Their work targeted a different architecture, replacing RAM by resistive CAM or ternary CAM in NVDIMM rather than in storage. Adopting associative processing architectures such as Goodyear Aerospace's STARAN or MPP to processing-in-storage is also suggested in [9].

## 3. PRINS Architecture

PRINS employs resistive memories, organized in RCAM modules, as described here. Resistive memories store information by modulating the resistance of nanoscale storage elements (memristors). They are nonvolatile, free of leakage power, and emerge as potential alternatives to charge-based memories, including NAND flash.

Memristors are two-terminal devices, where the resistance of the device is changed by the electrical current or voltage. The resistance of the memristor is bounded by a minimum resistance $R_{ON}$ (low resistive state, logic '1') and a maximum resistance $R_{OFF}$ (high resistive state, logic '0').

The metal-oxide resistive random access memory (RRAM) is one of the leading candidates for next-generation nonvolatile storage [4]. Its main features are high endurance and fast access speed. A test-chip of 32GB device with two RRAM-based memory layers and a CMOS logic layer underneath has been demonstrated [49].

RCAM is a scalable and highly dense alternative to CMOS CAM. A number of resistive CAM and ternary CAM cell designs have been proposed [5][23][51][52][74]. Our PRINS architecture uses a resistive crossbar and additional peripheral circuitry (Figure 2) to support associative storage and processing.

### 3.1 RCAM module

RCAM module, presented in Figure 2, is the heart of PRINS architecture. It comprises a resistive memory crossbar, in which each memory line is also a baseline processing unit (PU), and peripheral circuitry. The latter includes key and mask registers, tag logic, and two optional circuits: a tag counter, or reduction tree, and a daisy-chain interconnect. The basic RCAM cell is created by virtually pairing two RRAM cells (memristors), holding complementary values $R$ and $\bar{R}$.

The *key* register (Figure 2a) contains a key data word to be written or compared against. The *mask* register defines the active fields for write, compare and read operations, enabling bit selectivity. The *tag* marks the rows that are matched by the compare operation and are to be affected by the successive parallel write. A daisy-chain like bitwise interconnect allows PUs to intercommunicate, all PUs in parallel. The tag counter is a reduction (adding) tree, enabling logarithmic summation of tag bits. This operation is useful whenever a vector needs to be reduced to a scalar.

RCAM compare operation is performed as follows. The Match/Word line is precharged and the key is set on Bit and Bit-not lines. In the columns that are ignored during comparison, the Bit and Bit-not lines are kept floating. If all unmasked bits in a row match the key (*i.e.*, when Bit line '1' is applied to an $R_{ON}$ memristor and Bit-not line '0' is applied to an $R_{OFF}$ memristor, or vice versa), the Match/Word line remains high, and '1' is sampled into the corresponding tag bit. If at least one bit is mismatched, the Match/Word line discharges through an $R_{ON}$ memristor and '0' is sampled into the tag.

Write operation is performed in two phases. First, the $V \geq V_{ON}$ voltage (where $V_{ON}$ is a threshold voltage required to switch to the "on" state) is asserted to applicable Bit lines (to write '1's) and Bit-not lines (to write '0's). Second, the $V \leq$

$V_{OFF}$ voltage (where $V_{OFF}$ is a threshold voltage to switch to the "off" state) is asserted to Bit-not lines (to complement the '1's) and Bit lines (to complement '0's). The write affects only the tagged rows.

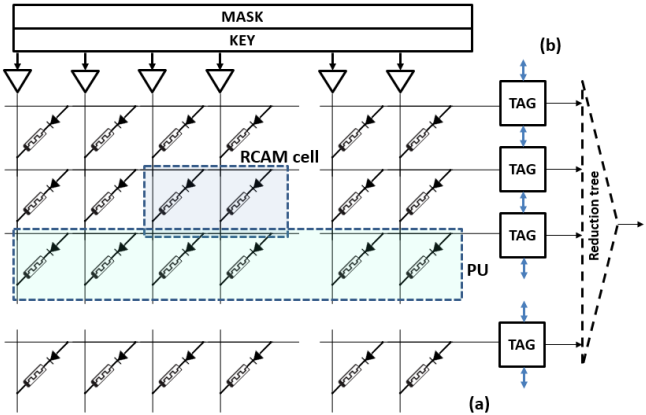

**Figure 2: RCAM Module: (a) Resistive Crossbar and (b) Peripheral Circuitry.**

Memristor sub-nanosecond switching time [67][71] allows GHz PRINS operation. The energy consumption during compare may be less than 1fJ per bit. The write energy is in the 100fJ per bit range [67][73][75], which may be prohibitively high for simultaneous parallel writing of the entire RCAM storage; the energy consumption is addressed in Section 6.

Another factor which potentially limits PRINS is endurance (the number of program/write cycles that can be applied to a memristor before it becomes unreliable). Resistive memory endurance is shown at about $10^{12}$ [67][75], which may suffice for only about one month. However, studies predict that the endurance of resistive memories may grow to the $10^{14} - 10^{15}$ range [23][58], extending PRINS endurance to a number of years.

### 3.2 Tag and Match Circuits

The tag logic is presented in Figure 3. It comprises a pre-charge circuit, a Match line sense amplifier, a tag latch, a multiplexor (part of the daisy-chain tag connectivity), a first_match circuit and an if_match circuit. The Match line is pre-charged during compare. The tag latch holds the result of compare. The First_match circuit implements 'first_match', a frequent associative operation, by keeping only the first matching tag and resetting the remaining tags (Section 5.2). If_match, another frequent associative operation, returns '1' if compare operation results in at least one match.

### 3.3 System Architecture and PRINS scaling

Conceptually, PRINS may comprise hundreds of millions or even billions of rows (PUs). Due to timing and thermal considerations, PRINS is divided into multiple daisy-chained RCAM modules (Figure 4), possibly implemented by separate ICs.

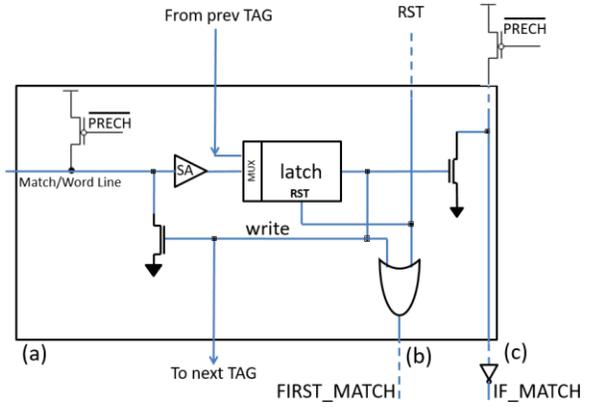

**Figure 3. Tag Logic: (a) Tag, (b) First_match, (c) If_match.**

PRINS controller is responsible for managing the processing operations of PRINS. It issues instructions, sets the key and mask registers, handles control sequences and executes read requests. In addition, it contains a data buffer, which stores the reduction tree outputs. The controller may also perform some baseline processing, such as normalization of the reduction tree results. Storage management unit orchestrates the storage operations, controlling read, write, translation, logical block mapping, wear leveling, etc.

The scaling of conventional near-data processing architectures may be limited, similarly to parallel manycore von Neumann architectures. When growing internal bandwidth of the storage arrays is met by increasing number of in-storage processing cores, the storage array to in-storage processor communication bottleneck worsens. As a result, the performance of a conventional processing-in-storage system may saturate or even diminish.

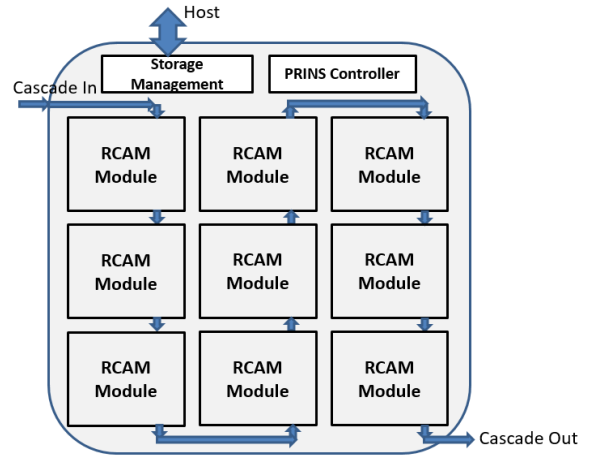

**Figure 4: PRINS is composed of multiple daisy-chained RCAM modules (potentially placed in separate ICs) (the figure is not to scale).**

PRINS is relatively simple to scale just by cascading RCAM modules. Daisy-chain connectivity allows an easy partitioning of PRINS into separate ICs. The inherent parallelism of PRINS allows increasing the performance of many workloads almost linearly as the datasets grow along with storage size. Since the bulk of data is never transferred outside the storage arrays through a bandwidth-limited communication interface, the performance limit is pushed further away.

### 3.4 PRINS position in memory hierarchy

The top-level view of PRINS and its possible positions within the memory hierarchy are presented in Figure 5.

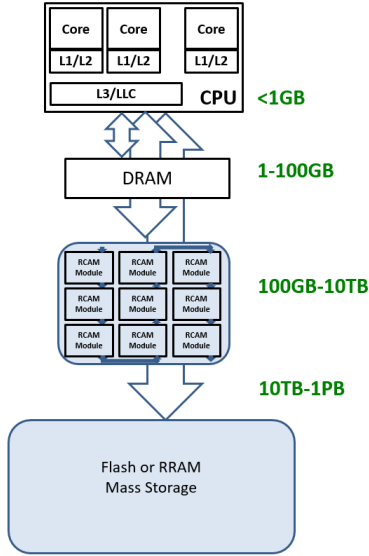

**Figure 5: PRINS as an intermediate level (storage class memory) (the figure is not to scale). Possible memory size ranges are shown on the right.**

PRINS can be implemented as an intermediate level in the memory hierarchy, between the main memory and mass storage, similar to a storage class memory, as shown in Figure 5. Such an option could provide a better performance/cost tradeoff.

### 4. Associative Processing

Associative processor is based on CAM, which allows comparing the stored data words to a search key pattern, tagging the matching words, and writing another key pattern to all tagged words. Associative processor is a non-von-Neumann in-memory computer. It performs no computations in conventional sense. Most arithmetic and logic operations can be structured as series of Boolean functions, which are implemented by associative processor using truth table-like execution.

The dataset is stored in the RCAM, typically one data element per RCAM row (PU). The truth table entries, embedded in the PRINS microcode, are broadcast entry-by-entry by the PRINS controller. The input section of each truth table entry is matched against the entire RCAM content (the entire dataset). The matching RCAM rows are tagged, and the corresponding truth table output values are written into the designated fields of the tagged rows. For an $m$-bit argument $x$ ($x \in$ dataset), any Boolean function $b(x)$ has $2^m$ possible output values. Therefore, a naïve associative computing operation would incur $O(2^m)$ cycles, regardless of the dataset size.

More efficiently, arithmetic operations can be performed in PRINS in a word-parallel, bit-serial manner, reducing time complexity from $O(2^m)$ to $O(m)$. For instance, vector addition may be performed as follows [25]. Suppose that two $m$-bit RCAM columns hold vectors A and B; the sum S=A+B is written onto another $m$-bit column S (Figure 6). A one-bit column C holds the carry bit. The operation is carried out as $m$ single-bit additions (2):

$$c[:] \mid s[:]_i = a[:]_i + b[:]_i + c[:], \quad i = 0, \ldots, m-1 \quad (2)$$

where $i$ is the bit index, ':' means all elements of the vector, and $c$ and $s$ are, respectively, the carry and sum bits. The single-bit addition is carried out in a series of steps. In each step, one entry of the single-bit (full) adder truth table (a three bit input pattern, Figure 6(a)) is matched against the contents of the $a[:]_i, b[:]_i, c[:]$ bit columns and the matching rows (PUs) are tagged; the logic result (two-bit output of the truth table, Figure 6(a)) is written into the $c[:]$ and $s_i[:]$ bits of all tagged rows. During that operation, all but three input bit columns and two output bit columns of RCAM are masked out in each step. Overall, eight steps of one compare and one write operation are performed to complete a single-bit addition over all RCAM rows (i.e. over all vector elements), regardless of the vectors A and B lengths.

A snapshot of such vector addition, for $m = 4$, zero bit of the vector elements and the 2nd entry of the truth table is shown in Figure 6. Figure 6(a) shows the truth table with the 2nd entry marked out. Figure 6(b) and (c) show compare and write operations respectively, against the backdrop of the RCAM map (with two 4-bit input vectors occupying bit-columns 0-3 and 4-7 respectively. The 4-bit output vector S is reserved columns 8-11, while bit column 12 is used for storing and updating the carry bit C.

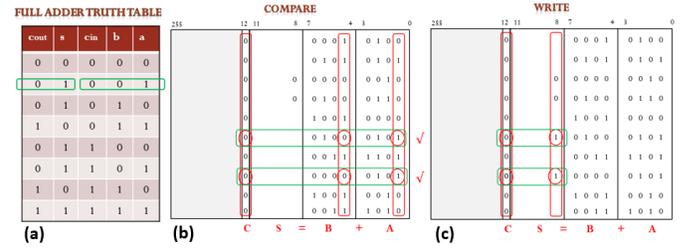

**Figure 6: Vector addition in RCAM example, for two 4-bit vectors A and B, snapshot at zero bit, 2nd entry of the truth table: (a) Full Adder Truth Table, (b) Compare, only c, $a_0$ and $b_0$ are affected, (c) Write, only c and $s_0$ in the tagged rows (PUs) are affected**

During compare (Figure 6(b)), the input pattern '001' is compared against bit columns c, $a_0$ and $b_0$, for all vector elements in parallel. The matching rows (two in this example) are tagged. During write (Figure 6(c)), the output pattern '01' is written in bit columns c and $s_0$ accordingly. Only the tagged rows are written. Each compare and write affect the entire dataset (vectors A, B and S).

A fixed-point $m$ bit addition and subtraction take $O(m)$ cycles. Fixed point multiplication and division in PRINS require $O(m^2)$ cycles. Single precision floating point multiplication takes 4,400 cycles [79], regardless of the dataset size.

### 5. Programming and applications

In this section, we briefly present PRINS data organization, instruction set and programming model. Afterwards, we discuss the implementation of several data-intensive workloads from different application fields, such as machine learning, data analytics, linear algebra and graph processing.

Somewhat less ambitious applications include basic data mining operations such as *grep* and *string matching,* addressed by several near-data processing-in-storage architectures [41][59]. Clearly, whereas the complexity of reading data out of storage arrays for performing search by in-SSD cores (or a host processor) is of linear time complexity, performing search in PRINS is closer to constant time complexity, and is not addressed in this work.

### 5.1 PRINS Data Organization

PRINS typically places one data element per RCAM row. Such data element normally occupies only a part of the row, while the rest of it is used for temporary storage. For example, in 128-bit wide row, data element may occupy bits 0 through 63, while bits 64 through 127 might be reserved for temporary variables, or used in associative arithmetic operations.

Data in associative memory is accessed by its content rather than its address. Data elements of the same dataset are normally identified by a unique index, or a class member ID. Unlike DRAM, RRAM and RCAM do not require a fixed, page-sized data allocation to operate efficiently. Therefore, individual data elements do not have to be placed in any specific or even dense order. They may rather be scattered in random sparse locations (rows) within the RCAM array, although typically in the same bit columns (fields).

### 5.2 PRINS Associative Instruction Set

In this section, we present the main PRINS associative instructions.

1. **compare (y1==x1, y2==x2, …., yn==xn)**. Compares the key x (stored in the key register) to the field y (masked by mask register) in the entire RCAM array.
2. **write (y1=x1, y2=x2, …, yn=xn)**. Writes the value x stored in the key register to the field y (masked by mask register) in the entire RCAM array. Write affects only the tagged RCAM rows (could be multiple rows).
3. **read (y)**. Reads field y (masked by mask register) from a tagged RCAM row to the key register.
4. **if_match**. Signals "1" if there is at least one match in the entire RCAM array.
5. **first_match**. Reset all tags set by compare but the first (top-most) one.

### 5.3 Programming PRINS

An external host may run an operating system and sequential code, and delegates processing-in-storage tasks to PRINS. PRINS implements these tasks as parallel SIMD kernels. We analyze applications to identify the highly parallelizable data intensive SIMD parts. We divide the application into sequential compute intensive parts (executed in the host) and massively parallel data intensive fractions (executed in PRINS). The code intended to run in PRINS is translated into associative primitives that are downloaded into and executed by the PRINS controller (Figure 4). Presently, PRINS code is manually encoded at assembly language level.

The host invokes PRINS to perform its code fraction. The host uses memory-mapped registers to communicate with PRINS through the host CPU - PRINS interface. The host transfers parameters such as data starting addresses, kernel configuration starting addresses, and kernel ID and triggers kernel execution/reconfiguration on PRINS by writing to the kernel registers. Once PRINS starts the execution, it can access the parameters from those registers. PRINS can notify the host of its execution status by writing to the status registers. The host periodically polls these memory-mapped status registers to check for kernel completion or exceptions. The status register read by the host does not intervene in PRINS operation. Based on the kernel running on PRINS and the dataset size of the kernel, the host can accurately estimate the execution time of a kernel, thereby polling status when the kernel execution is about to finish. Once PRINS execution completes, the host can access the PRINS output.

There is no hardware support for data coherence between the host CPU and PRINS. PRINS has no access to the host main memory or on-chip cache. Therefore, the datasets on which PRINS operates must reside in PRINS and should not be left in the host memory. To avoid inconsistencies between the PRINS and host memory, PRINS storage is inaccessible to the host during PRINS operation.

### 5.4 Applications

The focus of PRINS is large scale data intensive applications, i.e. applications with very high bandwidth requirements, and with datasets that do not fit in a typical main memory. We perform a case study on a number of data intensive workloads from four important big-data application domains, as follows.

#### 5.4.1 Machine Learning

Euclidean distance calculation is a frequent bottleneck in clustering algorithms, where the distances between each one of a number of cluster centers and each one of a multitude of multidimensional (multi-attribute) samples ($X$) are calculated iteratively.

Figure 7 presents a fully associative algorithm for Euclidean distance calculation in PRINS. The samples are assumed to be stored in PRINS, an attribute per RCAM row.

---

**Algorithm 1** Euclidean Distance

---

// $X$: the group of samples; samples are not required to be
// placed in any specific order prior to execution;
// $n$: number of centers
// Every $x \in X$ is stored in a separate RCAM row
// Each of the $n$ cluster centers: ($i_{center}$, $Center$)
1. For each $i_{center} \in [1, n]$: // $i_{center}$ is used as ID to //associatively mark the relevant samples
2.     Do-all $x \in X$: //at all RCAM rows in parallel
3.        Write $center$ coordinates to $temp$ column
4.        For each $attr \in \{sample\ attributes\}$:
5.           $dist_{attr} \leftarrow x_{attr} - center_{attr}$
6.           $sqEucDist_{attr} \leftarrow (dist_{attr})^2$ // Associative mult
7.           $sqEucDist_{to_{center}} \leftarrow sqEucDist_{to_{center}} + sqEucDist_{attr}$

**Figure 7: PRINS based Euclidean distance calculation**

All arithmetic operations (sub, add, square) are implemented associatively, in bit-serial manner, as series of compare and write commands. Lines 1-3 are performed in parallel for all attributes of all samples. Lines 4-7 are executed attribute by attribute, in parallel for all samples. The Euclidean distance calculation time does not depend on the number of samples, only on their dimensions.

Dot product calculation is a recurring bottleneck in classification algorithms, such as SVM, which iteratively calculates dot product between a hyperplane vector $H$ and a multitude of vectors $X$.

Figure 8 presents a fully associative algorithm for dot product calculation. The vectors are assumed to be stored in PRINS.

Arithmetic operations (multiply and add) are implemented associatively, in bit-serial manner, as series of compare and write commands. The For loop in line 1 is performed in parallel for all vectors. The dot product calculation time hence does not depend on the number of vectors, only on their size.

---

**Algorithm 2** Dot product

// $X$: input vectors of size $n$; Vectors are not required
// to be placed in any specific order prior to execution;
// $H$: Hyperplane vector
// Every $x \in X$ is stored in a separate RCAM row
1. For each $i \in [1, n]$: // $i$ is used as ID to //associatively mark the relevant elements
2.    Do-all $x \in X$: //at all RCAM rows in parallel
3.       $Mult_i \leftarrow x_i \times H_i$ // Associative mult
4.       $DP_i = DP_i + Mult_i$ // Associative add

**Figure 8: PRINS based dot product calculation**

### 5.4.2 Data Analytics

We calculate a $m$-bin histogram from 32-bit data (stored as a vector of size $n$). We further present a fully associative histogram calculation algorithm (Figure 9). We set $m = 256$ to allow a single-operation 1-byte shift to generate a bin index in an in-host implementation of histogram, used for comparison to PRINS implementation. The 32-bit samples are assumed to be stored in PRINS, a sample per RCAM row. Note that the histogram calculation makes use of the reduction tree.

---

**Algorithm 3** Histogram

// $X$: the group of samples; samples are not required
// to be placed in any specific order prior to
// execution; $m$: number of bins
// Every $x \in X$ is stored in a separate RCAM row
1. For each $i_{bin} \in [1, m]$: // $i_{bin}$ is used as ID to //associatively mark the relevant samples
2.    Do-all $x \in X$: //at all RCAM rows in parallel
3.       compare $i_{bin}$ to the bits [31…24] of $x$
4.       $H_{i_{bin}} \leftarrow Reduction(tagged\ rows)$

**Figure 9: PRINS based histogram calculation**

### 5.4.3 Linear Algebra

Sparse matrix by vector multiplication (SpMV) is a data intensive kernel widely used across the entire spectrum of machine learning algorithms. We propose a fully associative algorithm for SpMV execution in PRINS. Revised versions of this algorithm can be used for dense matrix multiplication and sparse matrix by matrix [79] multiplication.

Figure 10 presents the algorithm of PRINS SpMV. Matrix A is assumed to be stored in PRINS in Compressed Sparse Row (CSR) format, where each nonzero element $e_A$ is stored alongside its column index $i_A$.

The algorithm includes three parts. The first part, broadcast, consists of a loop going over the elements of vector $B$. In the first cycle, the index of an element of $B$, $i_B$, is compared against the column index field of the entire matrix $A$ (in parallel for all nonzero elements of $A$, using the compare command). All index-matching rows holding nonzero elements of matrix $A$ are tagged.

In the second cycle, B element $e_B$ is written simultaneously into all tagged rows, alongside the index-matched elements of matrix $A$. The loop is repeated for all elements of vector $B$. Upon completion, each nonzero pair of elements of $A$ and $B$ required to calculate the product vector C is aligned (stored in the same row) in the RCAM.

The second part (step 4) is the associative multiplication of the $e_A, e_B$ pairs, performed associatively in parallel for all pairs. The number of multiplications performed simultaneously equals the number of nonzero elements in $A$.

The third part sums the products along each row of $A$ (lines 5, 6) using the reduction tree.

PRINS SpMV has the computational complexity of $O(n_A)$ where $n_A$ is the matrix dimension (assuming square matrix for simplicity).

---

**Algorithm 4** SpMV

//Let A, B, C denote matrix A and vectors B and C.
//Each RCAM row holds a non-zero element of A ($e_A, i_A$)
// Matrix elements are not required to be placed in any
// specific order prior to execution;

   // Broadcast
1. For each $e_B \in \{elements\ of\ B\}$:
   // Compare $i_B$ with all column indices of A, $i_A$
2.    Compare $i_B$ to all $i_A$
   // Write $e_B$ into all matching rows
3.    Write $e_B$
   // Associatively multiply the entire A by B
4. $PR \leftarrow e_B * e_A$   // $PR$ is a matrix
   // Reduction: all rows of A in parallel, each row is tallied
5. For each (non-zero) row $k$ of $A$:
6.    $C_k \leftarrow Reduction(PR_k)$
   // $C$ has non-zero elements where $A$ has non-zero rows

**Figure 10: PRINS based SpMV pseudocode.**

### 5.4.4 Large Scale Graph Processing

Breadth first search (BFS) is an algorithm for traversing or searching graphs. BFS PRINS row mapping is presented in Table 2.

**Table 2. BFS Data Format and PRINS Row Mapping**

| Bits 0-47 | Bits 48-95 | Bit 96 | Bit 97 | Bits 98-145 | Bits 146-153 |
|---|---|---|---|---|---|
| Vertex ID | Successor ID | Visited Bit | Visited From Bit | Predecessor ID | Distance |

The pseudocode of PRINS serial implementation of BFS is presented in Figure 11.

**Algorithm 5 BFS**

// Source vertex has distance set to 0
// All visited_bit set to 0
// All visited_from_bit set to 0
1. j=-1;
2. j++;
3. do {
4. compare [distance == j, visited_from_bit == 0];
5. if (if_match==0) go to (2)
6. first_match;
7. write [**visited_from_bit** = 1]; // marks current vertex as "visited from"
8. read [**vertexID, successorID, visited_bit**]; // successors' visited_bit
9. do for all with **visited_bit**==0 {
10. compare [vertexID == successorID]
11. write [**distance** = j+1, **predecessorID** = vertexID, **visited_bit** = 1];   // updates all successors
12. } // do in (9)
13. } // do in (3)

**Figure 11: Serial BFS pseudocode**

## 6. Evaluation

### 6.1 Simulation Platform and Methodology

We assume that PRINS is implemented in 28nm technology. We simulate PRINS using the associative processor simulator [78], with operating frequency of 500MHz. We have developed an in-house power simulator to evaluate the power consumption of the PRINS. The latency and energy figures used by both the timing and power simulations are obtained using SPICE simulation with TEAM model [45].

The most logical baseline for the comparative analysis of PRINS performance and power efficiency is other data-centric processing architectures. However, such comparison is impractical, for the following reasons.

In some data centric designs, only relative (normalized) performance and energy results are reported. The lack of absolute performance and energy consumption figures makes the comparison impossible. Examples include graph processing in Tesseract [1], machine learning and in-memory data analytics in PIM-enabled instructions architecture [2], graph processing, histogram and deep neural network in 3D-NDP [53], graph processing and SpMV in TOP-PIM [80], a variety of workloads from various benchmarks such as SPLASH-2 and Rodinia in NDA [24], BFS and bitwise OR in Pinatubo [47], and several OpenCL applications in APIM [37].

When absolute performance and energy consumption figures are provided, e.g., in Intelligent SSD [8] for K-Means, not enough implementation details are disclosed for a meaningful comparison.

In processing-in-SSD architectures, the evaluation typically focuses on low performance applications (small scale or essentially sequential). Applications include query processing [59], heartbeat detection [10], and fgrep8 [19]. There is little merit in applying PRINS to such low performance workloads.

Another class of data centric works focuses on application-specific accelerators, such as analog memristive neural network accelerators PRIME [13] and ISAAC [66], and memristive Boltzmann machine [11]. We focus on programmable multi-application approach.

PRINS is a processing-in-storage architecture, capable of internally maintaining the entire dataset. The alternative is a computer architecture (either data centric or CPU centric) where the dataset does not fit in internal memory, therefore requiring an external storage. Such external storage could either be a SSD, or a NVDIMM based storage [34], or a dedicated storage appliance [35]. The bandwidth of such external storage is typically limited. For example, a high-end storage appliance may be limited by bandwidth of 10GB/s [35]. NVDIMM storage typically provides higher, although also limited bandwidth, for example 24GB/s [34].

The performance of such architecture is defined by the roofline model [72] as follows:

$$Attainable\ Perf = \min(Peak\ Perf,\ AI \times Peak\ Storage\ BW) \qquad (3)$$

where $Peak\ Perf$ is the peak theoretical performance of the computer architecture, $AI$ is arithmetic (or operational) intensity of a workload, and $Peak\ Storage\ BW$ is the peak external storage bandwidth (as demonstrated in Figure 15). In data intensive applications, characterized by low $AI$, the attainable performance of an architecture is likely to be limited by its peak storage bandwidth. Therefore, we use the term $AI \times Peak\ Storage\ BW$ as a baseline for our evaluation, presenting PRINS performance figures relative to it.

We simulate Euclidean distance calculation in PRINS using a number of synthetic vectors of sizes of 1M, 10M and 100M. Euclidean distance requires three floating point operations per each memory access (to fetch a vector attribute), assuming that center values and resulting Euclidean distance values are stored locally in the host (in its on-chip cache). Assuming single precision floating point, arithmetic intensity of Euclidean distance calculation is $AI = 3/4\ [\frac{FLOP}{B}]$. The attainable performance of Euclidean distance calculation is 7.5GFLOPS for a storage appliance and 18GFLOPS for a NVDIMM storage, although the peak theoretical performance of a reference computer architecture could be much higher. The Euclidean distance performance, normalized to the performance of a reference architecture with a bandwidth-limited external storage (either storage appliance or NVDIMM), is shown in Figure 12.

The power efficiency of PRINS Euclidean distance implementation is 2.9 GFLOPS/W.

We simulate dot product calculation in PRINS using synthetic 16-dimensional vectors, with the number of vectors of sizes of 1M, 10M and 100M. Dot product requires 2 FLOP per each memory access (to fetch a vector attribute), assuming that hyperplane vector and dot product values are stored locally in the host (in its on-chip cache). Assuming single precision floating point, arithmetic intensity of dot product calculation is $AI = 2/4\ [\frac{FLOP}{B}]$. The attainable performance of Dot product calculation is 5GFLOPS for a storage appliance and approximately 12GFLOPS for a NVDIMM storage. The dot product performance, normalized to the performance of a reference architecture with a bandwidth-limited external storage (either storage appliance or NVDIMM), is shown in Figure 12.

The power efficiency of PRINS dot product implementation is approximately 2.7 GFLOPS/W.

We simulate a 256-bin histogram calculation in PRINS using synthetic 32-bit integer vectors, with vector sizes of 1M, 10M and 100M. Histogram calculation requires 2 OP (a byte shift to generate a bin index and an increment) per each memory access (to fetch a 32-bit sample), assuming the results are stored locally (in the host on-chip cache). Arithmetic intensity of histogram calculation is $AI = 2/4\ [\frac{OP}{B}]$. The attainable performance of Euclidean distance calculation is 7.5GFLOPS for a storage appliance and approximately 18GFLOPS for a NVDIMM storage. The normalized histogram performance is shown in Figure 12.

The power efficiency of PRINS histogram implementation is 2.4 GFLOPS/W.

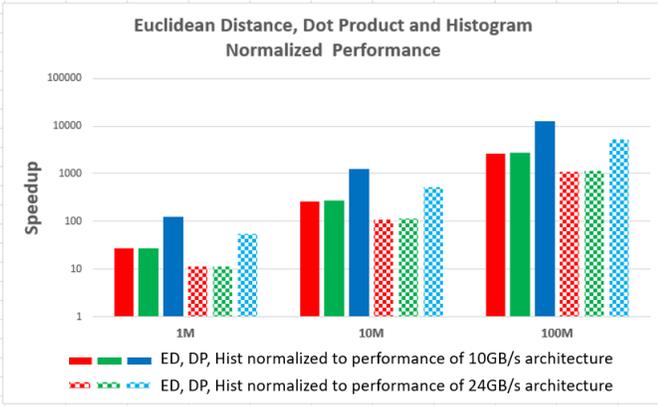

**Figure 12: PRINS Performance for Euclidean Distance (ED), Dot Product (DP) and Histogram (Hist), normalized to external storage bandwidth-limited (10GB/s and 24GB/s) architecture**

To simulate sparse matrix multiplication, we used 18 square matrices from the UFL Sparse Matrix Collection [17] (listed in Figure 13), having 1.2 through 29 million nonzero elements.

Assuming single precision floating point, arithmetic intensity of SpMV is $AI = 1/6\ [\frac{FLOP}{B}]$ [65]. The performance of SpMV, normalized to storage appliance and NVDIMM, is presented in Figure 13(a). The results are presented in the order of increasing matrix density, expressed as $\frac{nnz_A}{n_A}$ where $n_A$ is the matrix dimension and $nnz_A$ is the number of nonzero elements. PRINS relative performance grows with matrix density because all multiplications of vector elements by nonzero matrix elements are done in PRINS SpMV in parallel. The simulated power efficiency of the SpMV is presented in Figure 13(b). It ranges from 3 to 4GFLOPS/W.

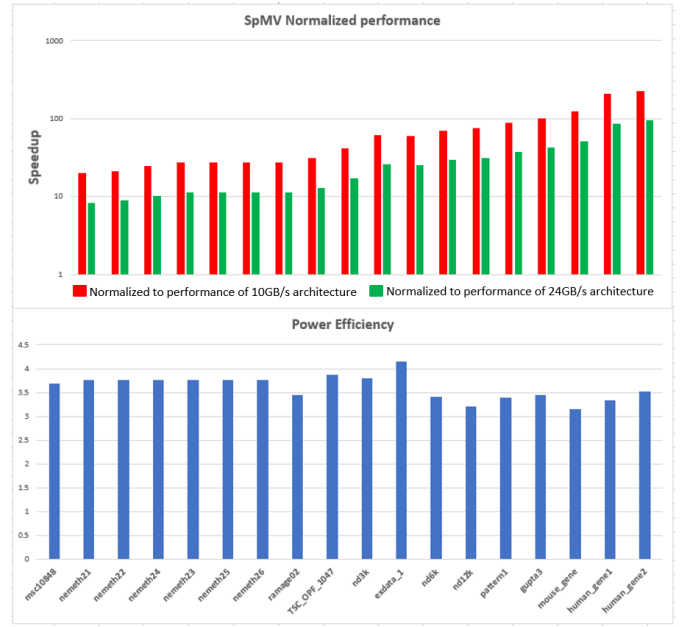

**Figure 13: SpMV (a) Normalized Performance, (b) Power Efficiency**

We simulate BFS using the graphs presented in Table 3. BFS requires two operations per two memory accesses (to fetch a vertex ID and save a new distance value and the predecessor ID). Arithmetic intensity of BFS is $AI = 1/4\ [\frac{OP}{B}]$. We measure BFS performance in Traversed Edges Per Second (TEPS). The attainable performance of BFS is 2.5GTEPS for a storage appliance and approximately 6GTEPS for a NVDIMM storage, although the peak theoretical performance of a reference computer architecture could be much higher. PRINS BFS normalized performance is presented in Figure 14. The results are presented in the order of increasing average out-degree (Avg D).

**Table 3. Graphs Used in Evaluation**

| Graph | V [M] | E [M] | Avg D | Max D |
|---|---|---|---|---|
| indochina-2004 | 5.3 | 79 | 15 | 19409 |
| arabic-2005 | 23 | 640 | 28 | 575,618 |
| it-2004 | 41 | 1151 | 28 | 1,326,745 |
| sk-2005 | 50.6 | 1,949 | 38 | 8,563,808 |
| kron_g500-logn21 | 2.1 | 182 | 87 | 213,905 |
| hollywood-09 | 1.1 | 114 | 100 | 11,468 |

V-vertices, E-edges, D- out-degree; Vertices and edges are in [Millions]

For all workloads examined here, PRINS performance is limited by the density of the problem, rather than by memory bandwidth. Euclidean distance and dot product are examples of dense problems (the entire dataset is processed simultaneously), where PRINS exhibits the best performance. In SpMV, performance is a function of matrix density. The denser the matrix (meaning more multiply-accumulate operations simultaneously), the better the performance. In BFS, PRINS achieves only up to 7 times better performance than external storage bandwidth-limited architectures, due to serial implementation (vertices are examined serially and speedup is limited by the average out-degree of the graph).

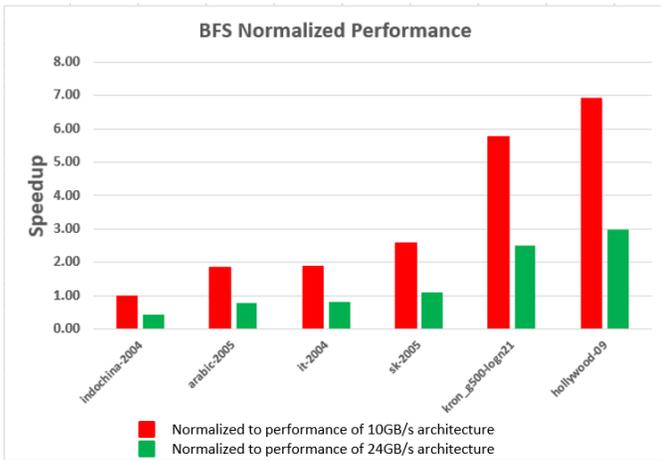

**Figure 14. BFS Normalized Performance**

The peak potential of PRINS (with 4TB of storage in this example) is illustrated using the Roofline model in Figure 15. It shows the roofline model of PRINS against the backdrop of KNL [20], to which we add a chart accounting for an external storage appliance access. Since PRINS requires no external access, its attainable performance is only limited by its ultra-high internal bandwidth. For instance, peak internal bandwidth is attained on a transfer of an entire bit column to the tag register. Another example is the broadcast of a single data item to the entire storage (e.g., the broadcast in SpMV, Figure 10). The peak theoretical performance is calculated using a single precision floating point multiply-accumulate operation, performed in parallel on the entire dataset (assuming the dataset matches the PRINS size, i.e. 1T 32bit data elements).

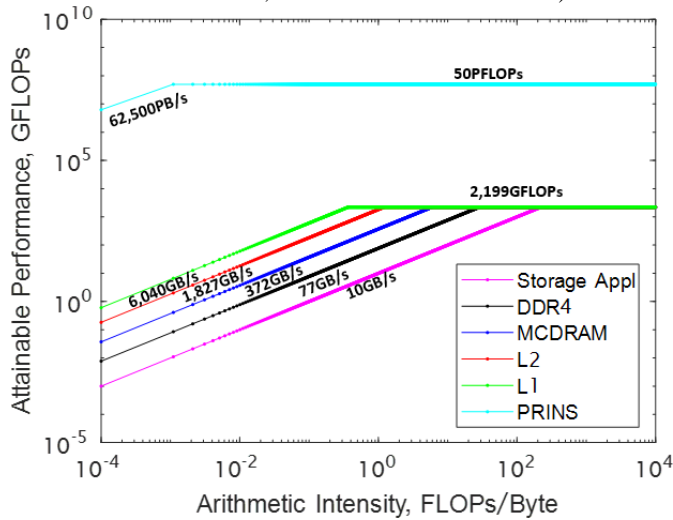

**Figure 15: Roofline model based on [20], amended by BW chart of an external storage appliance and a model for 4TB PRINS.**

## 7. Conclusions

Near-data processing-in-storage is inherently limited because it is based on replicating von Neumann architecture near storage. Therefore, it potentially faces some of von Neumann architecture problems, such as the bandwidth wall. To resolve this problem and allow for full utilization of ultra-high internal bandwidth of future resistive memory based storage, we propose PRINS, a novel in-data processing-in-storage architecture based on Resistive Content Addressable Memory (RCAM). Unlike near-data processing-in-storage, PRINS enables storage with *in-data* associative processing capabilities. It can contain billions to trillions of data rows, each row serving as an associative processing unit. PRINS requires no in-storage processing cores external to the storage arrays. There is no data transfer outside the storage arrays. Therefore, the internal bandwidth of the resistive memory based storage can be utilized to its fullest extent, considerably improving computation throughput of processing-in-storage system.

PRINS, capable of general purpose associative processing, has been applied to a variety of challenging data intensive problems in data analytics, machine learning and graph processing. The paper investigated Euclidian distance, dot product, histogram, Sparse Matrix-Vector multiplication and BFS and performed performance and power efficiency analysis of PRINS.

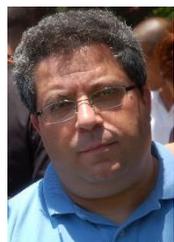

**Leonid Yavits** received his MSc and PhD in Electrical Engineering from the Technion. After graduating, he co-founded VisionTech where he co-designed a single chip MPEG2 codec. Following VisionTech's acquisition by Broadcom, he co-founded Horizon Semiconductors where he co-designed a Set Top Box on chip for cable and satellite TV. Leonid is a postdoc fellow in Electrical Engineering in the Technion. He co-authored a number of patents and research papers on SoC and ASIC. His research interests include non von Neumann computer architectures and processing in memory.

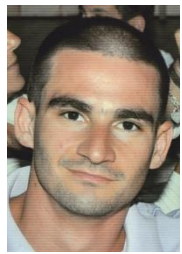

**Roman Kaplan** (M'17) received his BSC and MSc from the faculty of Electrical Engineering, Technion, Israel in 2009 and 2015, respectively. He is now a PhD candidate in the same faculty under the supervision of Prof. Ran Ginosar. Kaplan\s research interests are parallel computer architectures, in-data processing, accelerators for machine learning and big data, and novel computer architectures for bioinformatics applications.

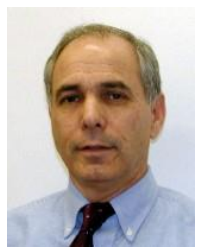

**Ran Ginosar** (M'78) received his BSc from the Technion—Israel Institute of Technology in 1978 (summa cum laude) and his PhD from Princeton University, USA, in 1982, both in Electrical and Computer Engineering. His Ph.D. research focused on shared-memory multiprocessors. He worked at AT&T Bell Laboratories in 1982-1983, and joined the Technion faculty in 1983. He was a visiting Associate Professor with the University of Utah in 1989-1990, and a visiting faculty with Intel Research Labs in 1997-1999. He is a Professor at the Department of Electrical Engineering and serves as Head of the VLSI Systems Research Center at the Technion. His research interests include VLSI architecture, manycore computers, asynchronous logic and synchronization, networks on chip and biologic implant chips. He has co-founded several companies in various areas of VLSI systems.